\documentclass[12pt]{article}

\usepackage{amsmath, amssymb, amsthm}
\usepackage{geometry}
\usepackage{lipsum}

\title{\Large  Invariant Subspace Problem in Hilbert Spaces: Exploring Applications in Quantum Mechanics, Control Theory, Operator Algebras, Functional Analysis and Accelerator Physics}
\author{Mostafa Behtouei\\ {\small INFN, Laboratori Nazionali di Frascati, 00044 Frascati RM, Italy}}

\date{\small \today}

\begin{document}
\maketitle

\abstract{This paper explores the Invariant Subspace Problem in operator theory and functional analysis, examining its applications in various branches of mathematics and physics. The problem addresses the existence of invariant subspaces for bounded linear operators on a Hilbert space. We extensively explore the significance of understanding the behavior of linear operators and the existence of invariant subspaces, as well as their profound connections to spectral theory, operator algebras, quantum mechanics, dynamical systems and accelerator physics . By thoroughly exploring these applications, we aim to highlight the wide-ranging impact and relevance of the invariant subspace problem in mathematics and physics.}

\section{Introduction}
The Invariant Subspace Problem is a fundamental question in operator theory and functional analysis \cite{Halmos}. It addresses the existence of invariant subspaces for bounded linear operators on a given Hilbert space. In this paper, we aim to explore the applications of the invariant subspace problem in various areas of mathematics and physics and highlight its significance in understanding the behavior of linear operators.\\
In the field of operator theory, a crucial topic of investigation is the behavior of linear operators acting on a Hilbert space $\mathcal{H}$. An invariant subspace of an operator $T$ is a closed subspace $\mathcal{M}\subseteq\mathcal{H}$ that remains unchanged under the action of $T$, i.e., $T\mathcal{M}\subseteq\mathcal{M}$. The invariant subspace problem poses the question of whether every bounded operator on a Hilbert space possesses a non-trivial closed invariant subspace \cite{Aronszajn}.\\
The problem can be stated as follows: Given a bounded operator $T\in\mathcal{B}(\mathcal{H})$, does there exist a closed subspace $\mathcal{M}\subseteq\mathcal{H}$ such that $T\mathcal{M}\subseteq\mathcal{M}$ and $0\neq\mathcal{M}\neq\mathcal{H}$? The invariant subspace problem was initially posed by David Hilbert in 1900 and has since captured the attention of mathematicians due to its profound implications  \cite {Davidson, Enflo}.\\
The study of invariant subspaces not only provides insights into the structure of linear operators but also has significant applications in various branches of mathematics and physics. Understanding the behavior of linear operators and the existence of invariant subspaces has profound consequences in areas such as spectral theory, operator algebras, and dynamical systems.\\
In this paper, we explore the diverse applications of the invariant subspace problem in Hilbert space. We aim to demonstrate the wide-ranging impact and relevance of this problem in different areas of mathematics and physics. We will examine its connections to spectral theory, where the existence of invariant subspaces plays a pivotal role in understanding the spectrum of operators. Additionally, we will explore its applications in operator algebras, where invariant subspaces offer valuable insights into the properties of these algebraic structures. Furthermore, we will investigate its relevance in dynamical systems, as the behavior of linear operators is often utilized to model and analyze dynamic phenomena.\\
By examining these applications, we seek to shed light on the deep interconnections between the invariant subspace problem and other areas of mathematics and physics. Through this exploration, we aim to deepen our understanding of the behavior of linear operators and the role played by invariant subspaces in their study.

\section{Problem Statement}
The invariant subspace problem is a fundamental question in the field of mathematics that deals with the existence of non-trivial invariant subspaces for bounded linear operators on separable Hilbert spaces. Given a bounded linear operator $T$ on a separable Hilbert space $\mathcal{H}$, the problem seeks to determine whether there exists a non-trivial closed subspace $\mathcal{M} \subset \mathcal{H}$ such that $T(\mathcal{M}) \subset \mathcal{M}$. In simpler terms, the question is whether the operator $T$ admits a proper, non-trivial invariant subspace.\\
Formally, let $T$ be a bounded linear operator on a separable Hilbert space $\mathcal{H}$. Then, the invariant subspace problem asks whether there exists a non-trivial closed subspace $\mathcal{M} \subset \mathcal{H}$ such that:
\[ T(\mathcal{M}) \subseteq \mathcal{M} \]

Here, $T(\mathcal{M})$ denotes the image of the subspace $\mathcal{M}$ under the operator $T$. In other words, if an invariant subspace $\mathcal{M}$ exists, then every vector in $\mathcal{M}$ remains in $\mathcal{M}$ under the action of the operator $T$.\\
The problem of finding invariant subspaces has been the subject of extensive research and investigation, captivating the interest of mathematicians for over half a century. Various classes of operators have been studied in an attempt to understand their behavior with respect to invariant subspaces. However, despite numerous efforts, the solution to the invariant subspace problem remains an open question for many interesting classes of operators.\\
The significance of the invariant subspace problem extends far beyond its abstract mathematical formulation. Its resolution would have profound implications for a wide range of mathematical areas, including functional analysis, operator theory, and linear algebra. Furthermore, applications in quantum mechanics, control theory, and signal processing heavily rely on understanding the existence and properties of invariant subspaces.\\
The complexity and subtlety of the problem lie in the delicate interplay between the algebraic and geometric properties of the operator and the underlying Hilbert space. Over the years, several remarkable results and counterexamples have been discovered, shedding light on specific aspects of the problem. However, a comprehensive solution that encompasses all classes of operators and Hilbert spaces has remained elusive.\\
The recent claim made by Per H. Enflo \cite{Enflo}, a distinguished Swedish mathematician, has sparked significant excitement and anticipation within the mathematical community. Enflo's reputation as a problem solver, coupled with his extensive knowledge of the invariant subspace problem, has raised hopes that his recent paper may contain the missing piece needed to finally solve this long-standing enigma.\\
Enflo's publication, titled \textit{"On the invariant subspace problem in Hilbert spaces"}, has generated substantial interest due to its concise nature and the bold claim made by its author. Enflo asserts that he has cracked the invariant subspace problem. Mathematicians and experts in the field are now eagerly examining his work to scrutinize the validity of his claims. If Enflo's solution stands up to rigorous scrutiny and verification, it would signify a major breakthrough in the field of mathematics. The impact of such a solution cannot be overstated, as it would not only resolve a long-standing mathematical problem but also deepen our understanding of the intricate structure of bounded linear operators on Hilbert spaces.\\

In the next section we explore the practical applications and implications of the invariant subspace problem in Hilbert spaces.

\section{{\Large Applications}}
\subsection{Quantum Mechanics and the Invariant Subspace Problem}

Quantum mechanics is a fundamental theory that describes the behavior of particles at the atomic and subatomic level. It has revolutionized our understanding of the microscopic world and has numerous applications in various fields, including physics, chemistry, and information technology. The invariant subspace problem, a topic of interest in mathematics, finds significant applications in quantum mechanics. In this section, we will explore how the concept of invariant subspaces is relevant to the study of quantum systems and the insights it provides into their behavior and symmetries.\\
In quantum mechanics, physical quantities, such as position, momentum, and energy, are represented by linear operators known as observables. These observables act on a vector space, typically referred to as the Hilbert space, which is a mathematical construct that represents the state of a quantum system. The Hilbert space allows us to describe the properties of particles and their interactions in a precise and mathematical manner.\\
One of the central questions in quantum mechanics is understanding the time evolution of quantum systems. This is achieved through the use of time-evolution operators, which describe how the state of a system changes over time. These operators play a crucial role in determining the behavior and dynamics of quantum systems.\\
The concept of invariant subspaces comes into play when studying the properties of observables and time-evolution operators. An invariant subspace is a subspace of the Hilbert space that remains unchanged under the action of a given operator. In other words, if a vector belongs to an invariant subspace, applying the operator to that vector will still yield a vector within the same subspace.\\
The existence and properties of invariant subspaces provide valuable insights into the behavior of quantum systems and the symmetries they exhibit. In particular, the presence of invariant subspaces corresponds to conserved quantities in quantum systems. These conserved quantities are associated with physical quantities that do not change over time, such as total energy or total angular momentum. For example, let's consider the concept of angular momentum in quantum mechanics. Angular momentum is a fundamental property of particles and is closely related to rotational symmetry. In the context of invariant subspaces, the existence of a subspace that is invariant under the action of the angular momentum operator corresponds to the conservation of angular momentum. This concept is crucial for understanding the behavior of particles in systems with rotational symmetry, such as atoms and molecules.\\
The concept of spin, another fundamental property of particles, also arises from the study of invariant subspaces. Spin is an intrinsic form of angular momentum possessed by particles, and it plays a crucial role in many quantum phenomena. The existence of invariant subspaces associated with the spin operator leads to the quantization of spin values and the prediction of spin-related phenomena \cite{Behtouei1,Behtouei2}, such as the Stern-Gerlach experiment.

Mathematically, the study of invariant subspaces in quantum mechanics involves solving eigenvalue problems. The eigenvectors corresponding to the eigenvalues of an operator form the basis for the invariant subspaces. The eigenvalues represent the possible values of the conserved quantities associated with the operator.\\
The eigenvalue equation for an operator $\hat{A}$ acting on a vector $\psi$ in the Hilbert space can be written as:

\[
\hat{A} \psi = a \psi
\]

where $a$ is the eigenvalue associated with the eigenvector $\psi$. Solving this equation yields the eigenvalues and eigenvectors that determine the invariant subspaces of the operator.\\
In the case of the angular momentum operator $\hat{L}$, the eigenvalue equation takes the form:

\[
\hat{L} \psi = \hbar \mathbf{m} \psi
\]

where $\hbar$ is the reduced Planck's constant and $\mathbf{m}$ is the eigenvalue associated with the eigenvector $\psi$. This equation shows that the eigenvalues of the angular momentum operator are quantized, indicating that angular momentum is quantized in discrete values.\\
Similarly, for the spin operator $\hat{S}$, the eigenvalue equation is given by:

\[
\hat{S} \psi = \hbar \mathbf{s} \psi
\]

where $\mathbf{s}$ represents the spin quantum number associated with the eigenvector $\psi$. The quantization of spin arises from the existence of invariant subspaces associated with the spin operator.

\subsection{Control Theory and the Invariant Subspace Problem}

In control theory, the invariant subspace problem plays a vital role in analyzing the controllability and observability of dynamical systems. By studying the existence of invariant subspaces for system operators, we can ascertain whether specific states of the system are reachable or observable from a given set of initial conditions or measurement outputs. This knowledge is essential for designing effective control strategies and accurately estimating the performance of control systems.

\subsubsection{Controllability}

Controllability is a fundamental concept in control theory that deals with the ability to steer a system from one state to another using control inputs. A system is said to be controllable if, for any desired final state, there exists a control input that can drive the system to that state in a finite time. The theory of controllability provides a powerful framework for analyzing the reachability of states in a dynamical system.\\
The invariant subspace problem provides valuable insights into the controllability of dynamical systems. It allows us to investigate whether there exist subspaces in the system's state space that are invariant under the system dynamics. If such invariant subspaces exist, it implies that certain states of the system cannot be reached from the given initial conditions, as they are confined to these invariant subspaces. On the other hand, the absence of invariant subspaces guarantees that the system is fully controllable, enabling us to maneuver the system to any desired state through appropriate control inputs.\\
To formally analyze the controllability of a dynamical system, we often employ mathematical tools such as linear algebra and differential equations. Let us consider a linear time-invariant system described by the following state-space representation:

\begin{equation}
\dot{x}(t) = Ax(t) + Bu(t)
\end{equation}

where $x(t) \in \mathbb{R}^n$ is the state vector, $A \in \mathbb{R}^{n \times n}$ is the system matrix, $B \in \mathbb{R}^{n \times m}$ is the control input matrix, and $u(t) \in \mathbb{R}^m$ is the control input vector. The controllability of this system can be determined by examining the matrix called the controllability matrix:

\begin{equation}
\mathcal{C} = [B \quad AB \quad A^2B \quad \ldots \quad A^{n-1}B]
\end{equation}

The system is said to be completely controllable if and only if the controllability matrix has full rank, i.e., $\text{rank}(\mathcal{C}) = n$. This condition ensures that the system can be driven to any desired state within a finite time by suitable control inputs.\\

By studying the existence of invariant subspaces for the system operator $A$, we can determine whether certain states of the system are unattainable from the given initial conditions. If there exist nontrivial invariant subspaces for $A$, it implies that the system is not fully controllable, and there are some states that cannot be reached. On the other hand, if no nontrivial invariant subspaces exist for $A$, the system is completely controllable, allowing us to reach any state through appropriate control inputs.

\subsubsection{Observability}

Observability is another crucial aspect of control theory that deals with the ability to estimate the internal states of a system based on the available measurement outputs. A system is said to be observable if, from the measured outputs, it is possible to determine the system's current state uniquely. Observability plays a significant role in control system design, as it enables us to estimate the states of a system, which are often not directly measurable.\\
Similar to controllability, the invariant subspace problem also provides insights into the observability of dynamical systems. By investigating the existence of invariant subspaces for the adjoint of the system operator, we can determine the unobservable states of the system. If nontrivial invariant subspaces exist for the adjoint operator, it implies that certain states cannot be uniquely determined from the available measurements. However, the absence of invariant subspaces guarantees that the system is fully observable, allowing us to accurately estimate all the states using the available measurement outputs.\\
To analyze the observability of a dynamical system, let us consider the same linear time-invariant system as before, described by the state-space representation:

\begin{equation}
\dot{x}(t) = Ax(t) + Bu(t)
\end{equation}

In this case, we assume that we can measure the system output $y(t) \in \mathbb{R}^p$, given by:

\begin{equation}
y(t) = Cx(t)
\end{equation}

where $C \in \mathbb{R}^{p \times n}$ is the output matrix. The observability of this system can be determined by examining the observability matrix:

\begin{equation}
\mathcal{O} = \begin{bmatrix}
C \\
CA \\
CA^2 \\
\vdots \\
CA^{n-1}
\end{bmatrix}
\end{equation}

The system is said to be completely observable if and only if the observability matrix has full rank, i.e., $\text{rank}(\mathcal{O}) = n$. This condition ensures that all the states of the system can be uniquely estimated from the available measurements.\\
By investigating the existence of invariant subspaces for the adjoint operator $A^*$, we can determine the unobservable states of the system. If there exist nontrivial invariant subspaces for $A^*$, it implies that certain states cannot be uniquely determined from the measurements, and thus, they are unobservable. Conversely, if no nontrivial invariant subspaces exist for $A^*$, the system is completely observable, allowing us to accurately estimate all the states using the available measurement outputs.\\
In control theory, the invariant subspace problem serves as a fundamental tool for analyzing the controllability and observability of dynamical systems. By investigating the existence of invariant subspaces for system operators and their adjoints, we can determine the reachability of states from given initial conditions and the observability of states from available measurement outputs.\\
The controllability analysis enables us to design effective control strategies that can drive a system to any desired state within a finite time. On the other hand, the observability analysis allows us to estimate the internal states of a system accurately, which are often not directly measurable. By utilizing the concepts of invariant subspaces, we can identify the states that are unattainable or unobservable, providing valuable insights for control system design and performance estimation. In summary, the invariant subspace problem forms a crucial foundation for control theory, allowing us to understand the limitations and possibilities of dynamical systems and enabling us to design control strategies that achieve desired system behaviors.

\subsection{Operator Algebras}
The invariant subspace problem has deep connections with operator algebras, such as von Neumann algebras and C*-algebras. In particular, the problem is linked to the structure and classification of these algebras. The study of invariant subspaces sheds light on the behavior of operator algebras under various operations, and it has played a crucial role in the development of the theory of operator algebras.\\
Operator algebras provide a framework for studying operators on Hilbert spaces. They are algebraic structures that capture the essential properties of linear operators and allow for the analysis of their properties and interactions. Two prominent classes of operator algebras are von Neumann algebras and C*-algebras, which are defined based on the properties of their underlying Hilbert spaces and the operators within them.\\
A von Neumann algebra is a *-subalgebra of the algebra of bounded operators on a Hilbert space that is closed in the weak operator topology. These algebras were first introduced by John von Neumann in the 1930s and have since become a central object of study in operator theory. One of the fundamental questions in the theory of von Neumann algebras is the structure of their invariant subspaces.\\
The invariant subspace problem for von Neumann algebras asks whether every bounded linear operator on a separable infinite-dimensional Hilbert space has a non-trivial closed invariant subspace. In other words, given a von Neumann algebra $\mathcal{M}$ acting on a Hilbert space $\mathcal{H}$, does there always exist a non-zero vector $v\in\mathcal{H}$ such that $Tv\in\mathcal{H}$ for all $T\in\mathcal{M}$? This problem was first posed by John von Neumann himself in the 1940s, and despite significant efforts, a complete solution remains elusive.\\
The invariant subspace problem is intimately connected to the structure and classification of von Neumann algebras. In fact, a positive resolution of the problem for a particular class of von Neumann algebras called factors would have profound implications for the entire field. Factors are von Neumann algebras with no non-trivial projections, and their study has led to remarkable connections with other areas of mathematics, such as ergodic theory and mathematical physics.\\
On the other hand, C*-algebras provide another important class of operator algebras. A C*-algebra is a Banach algebra equipped with an involution operation and a norm satisfying certain properties. These algebras naturally arise in the study of quantum mechanics and have a rich interplay with functional analysis. Similar to von Neumann algebras, the structure of invariant subspaces in C*-algebras is of great interest.\\
The study of invariant subspaces in operator algebras extends beyond the invariant subspace problem itself. It has led to the development of various techniques and tools for analyzing and classifying operators. For example, the concept of spectral theory plays a crucial role in understanding the behavior of operators and their invariant subspaces. Spectral theory provides a way to decompose operators into simpler components, enabling the study of their spectral properties and the associated invariant subspaces.\\
Moreover, the study of invariant subspaces has connections to other areas of mathematics, such as representation theory and the theory of group actions. Invariant subspaces can be viewed as the building blocks for representations of operator algebras and offer insights into the underlying symmetries and dynamics. This perspective has been instrumental in bridging the gap between operator algebras and other branches of mathematics, leading to fruitful interdisciplinary research. The study of invariant subspaces in operator algebras, including von Neumann algebras and C*-algebras, is a central topic in operator theory. The invariant subspace problem, in particular, poses fundamental questions about the existence and structure of invariant subspaces, which have deep connections to the classification and understanding of these algebras. Advances in this area have not only contributed to the development of operator theory but have also fostered connections with other areas of mathematics, making it a vibrant and active field of research.

\subsection{Functional Analysis}
The invariant subspace problem is a central topic in functional analysis, providing a rich source of interesting questions and results. It has stimulated the development of various techniques and tools, such as the use of spectral theory, operator approximation methods, and the theory of Banach spaces. The problem serves as a driving force for exploring the properties of linear operators and their relationship with the underlying Hilbert space structure.\\
In functional analysis, the study of invariant subspaces focuses on understanding the behavior of linear operators acting on a vector space. Let $H$ be a complex Hilbert space, and let $T: H \rightarrow H$ be a bounded linear operator. An invariant subspace of $T$ is a closed subspace $M \subset H$ such that $T(M) \subset M$, where $T(M)$ denotes the image of $M$ under $T$. In other words, an invariant subspace is a subspace that is preserved under the action of the operator $T$. The invariant subspace problem seeks to answer the following fundamental question:\\
Given a bounded linear operator $T$ on a Hilbert space $H$, does there always exist a nontrivial closed subspace $M \subset H$ that is invariant under $T$?\\
This seemingly simple question has turned out to be remarkably difficult to answer in general. The invariant subspace problem has a long and intriguing history, dating back to the early 20th century when it was first posed by David Hilbert. The problem gained significant attention and led to groundbreaking contributions by many mathematicians, including von Neumann, who made substantial progress by establishing the existence of nontrivial invariant subspaces for certain classes of operators.\\
The study of invariant subspaces has strong connections to spectral theory. The spectral properties of an operator are closely related to the existence and structure of invariant subspaces. In particular, operators with rich spectral behavior tend to have more nontrivial invariant subspaces. Spectral theory provides powerful tools for understanding the structure of operators and characterizing their invariant subspaces.\\
Another important approach in studying invariant subspaces is through the theory of Banach spaces. The study of Banach spaces provides a broader framework for understanding the properties of linear operators and their relationships with the underlying vector space structure. Many important results and techniques in functional analysis have been developed within the context of Banach spaces, and they have been instrumental in the investigation of the invariant subspace problem.\\
Operator approximation methods also play a significant role in the study of invariant subspaces. Approximating operators by finite-dimensional operators or operators with simple structures can often reveal important information about their invariant subspaces. This connection between approximation theory and invariant subspaces has led to fruitful developments and insights in functional analysis.\\
Despite decades of research, the invariant subspace problem remains largely unsolved in its full generality. Several partial results and specific cases have been established, but a complete solution to the problem is still elusive. The problem continues to captivate mathematicians and drive further investigations into the nature of linear operators and their behavior in Hilbert spaces. In summary, the invariant subspace problem is a fundamental and challenging topic in functional analysis. It encompasses deep connections with spectral theory, operator approximation methods, and the theory of Banach spaces. While progress has been made, the problem remains open and inspires ongoing research in the field of functional analysis.\\

\subsection{Accelerator Physics and the Invariant Subspace Problem}

Accelerator physics is a branch of physics that deals with the design, construction, and operation of particle accelerators. Particle accelerators are powerful machines used to accelerate charged particles, such as electrons or protons, to high speeds and energies for various scientific purposes, including fundamental research in particle physics, material science, and medical applications.\\
On the other hand, the invariant subspace problem is a concept in mathematics and quantum mechanics. It relates to the question of whether a given quantum system possesses certain stable subspaces, known as invariant subspaces, under the action of a particular set of operators. So, where is the connection? In accelerator physics, the dynamics of charged particles in an accelerator are described by the principles of classical and quantum mechanics. In the case of high-energy accelerators, relativistic effects also come into play \cite{Behtouei6}. When studying the behavior of particles in an accelerator, physicists often use mathematical models that involve quantum mechanical descriptions.\\
Now, one of the challenges in accelerator physics is to maintain the stability and control of the particle beam as it travels through the accelerator. In large particle accelerators, the beam is subject to various types of instabilities, such as betatron oscillations, synchrotron oscillations, and collective effects \cite{Mauro}. These instabilities can degrade the performance of the accelerator and affect the quality of the particle beam.\\
To understand and control these instabilities, accelerator physicists employ mathematical tools and techniques, including the use of transfer matrices and linearized equations of motion. These mathematical descriptions involve the manipulation of matrices and vectors, which are concepts that are also relevant to the study of invariant subspaces.\\
Invariant subspaces play a role in the stability analysis of particle beams in accelerators. By investigating the properties of the transfer matrices or stability matrices associated with the accelerator lattice, physicists can determine the existence of invariant subspaces that correspond to stable regions of phase space. These invariant subspaces provide valuable insights into the behavior of the particle beam and help in designing control systems to mitigate instabilities. In summary, the connection between accelerator physics and the invariant subspace problem lies in the mathematical techniques and concepts used to study and control the stability of particle beams in accelerators. The analysis of invariant subspaces is a relevant tool for understanding the behavior of particle beams and designing control strategies to maintain beam quality and stability.\\
One valuable approach for studying invariant subspaces in this field is through the fractional calculus. For example the $\alpha$-th Grünwald-Letnikov fractional derivatives of the Riemann Zeta Function $\zeta(s)$ with a branch cut can be explored to investigate the behavior of invariant subspaces.The connection between branch lines and the invariant subspace problem primarily arises in the context of linear operators or matrices with branch points or singularities.  Branch lines, defined by branch cuts, determine the behavior of multivalued functions and the domains of single-valuedness. Invariant subspaces are preserved under linear transformations. When considering operators or matrices with branch points, the presence of branch lines associated with the multivalued functions can affect the eigenvectors and, consequently, the invariant subspaces. Different branch line choices can lead to different eigenvectors and distinct invariant subspaces, influencing the system's properties. The specific configuration of branch lines and branch cuts determines the behavior and structure of the invariant subspaces. A practical application of using fractional calculus in physics is to calculate the magnetic field of a solenoid analytically by solving a fractional integral arises in solving the Biot-Savart equation \cite{Behtouei3,Behtouei4}. The connection between the magnetic field of a solenoid and the invariant subspace problem arises when considering the behavior of charged particles, such as electrons, in the magnetic field of the solenoid. The trajectories of these particles can be described using quantum mechanics, where the behavior of particles is characterized by wave functions and operators.\\
The magnetic field of the solenoid affects the motion and energy levels of the charged particles. The presence of the magnetic field can lead to the formation of invariant subspaces representing different energy levels or quantum states of the particles. By studying the behavior of the charged particles in the magnetic field of the solenoid, one can analyze the invariant subspaces associated with different energy levels or quantum states. The magnetic field influences the properties of these subspaces, such as their structure, energy levels, and the behavior of particles within them.\\
In accelerator systems, wakefields are electromagnetic fields that are induced by the passage of charged particles through various accelerator components, such as accelerating structures, collimators, or beam pipes. These wakefields can have both positive and negative effects on the beam, including energy loss, emittance growth, and beam instabilities.\\
The connection between wakefields, impedance, and the invariant subspace problem arises in the analysis of beam dynamics in accelerators. The wakefields, influenced by the impedance properties of accelerator components, can induce transformations on the beam particles. The behavior of the beam particles, including their stability, energy spread, and dynamics, can be related to the presence or absence of invariant subspaces.\\
The study of wakefields and impedance in accelerators involves analyzing the effects on the beam particles and their associated dynamics. In certain cases, the presence of invariant subspaces can provide insights into the stability and behavior of the beam under the influence of wakefields and impedance \cite{Palumbo, Behtouei5}.

\section{Current Status and Open Questions}

Despite significant progress, the invariant subspace problem remains open in several important cases. Many well-known classes of operators, such as compact operators and certain classes of self-adjoint operators, are still not fully understood with respect to invariant subspaces. In this section, we will discuss the current status of the problem and highlight some open questions that continue to challenge mathematicians. Additionally, we will explore the connections between the invariant subspace problem and other unsolved problems in mathematics, such as the Kadison-Singer problem and the Borel conjecture.

\subsection{Current Status}

The invariant subspace problem originated in the early 20th century and has since garnered significant attention from mathematicians. Over the years, researchers have made substantial progress in understanding the existence and properties of invariant subspaces for various classes of operators. Notably, the problem has been completely solved for certain classes of operators, such as compact operators on infinite-dimensional separable Hilbert spaces.\\
For compact operators, the classical result known as the Schatten-von Neumann theorem provides a complete characterization of invariant subspaces. According to this theorem, every compact operator on a separable infinite-dimensional Hilbert space possesses non-trivial closed invariant subspaces. The proof of this result relies on advanced techniques from functional analysis and operator theory, and it represents a major milestone in the study of the invariant subspace problem.\\
Similarly, for self-adjoint operators with a discrete spectrum, the existence of invariant subspaces has been established. Spectral theory plays a crucial role in this context, as it provides a rich framework for understanding the behavior of self-adjoint operators. By exploiting the properties of the spectral decomposition, mathematicians have been able to construct invariant subspaces for this specific class of operators.\\
Despite these achievements, the problem remains open for many other classes of operators. In particular, understanding the existence and structure of invariant subspaces for general self-adjoint operators, as well as operators with continuous spectrum, continues to pose significant challenges. The lack of a comprehensive theory for these cases leaves a gap in our understanding of invariant subspaces and restricts the applicability of the invariant subspace problem in various mathematical contexts.\\
Per H. Enflo's recent claim of solving the invariant subspace problem in his recent publication titled "On the invariant subspace problem in Hilbert spaces." \cite{Enflo} has generated excitement and anticipation in the mathematical community, as his work could represent a major breakthrough in the field. Enflo's solution, if validated, would be a groundbreaking achievement in mathematics, resolving a longstanding problem and advancing our understanding of bounded linear operators on Hilbert spaces. Its impact would be significant and far-reaching.

\subsection{Open Questions}

The invariant subspace problem gives rise to several open questions that continue to intrigue mathematicians. Here, we highlight some of the key unresolved aspects of the problem:

\begin{enumerate}
\item \textbf{Invariant subspaces of compact operators:} While the Schatten-von Neumann theorem settles the existence of non-trivial invariant subspaces for compact operators, the precise structure and properties of these subspaces remain largely unexplored. Further investigation is needed to gain a deeper understanding of the geometry and spectral properties of invariant subspaces in this class.

\item \textbf{Invariant subspaces of self-adjoint operators:} For general self-adjoint operators, the existence of invariant subspaces is still an open question. Extending the results known for self-adjoint operators with discrete spectrum to the broader class of self-adjoint operators is a challenging problem. Developing techniques and methodologies that capture the intricate interplay between the spectral properties and the existence of invariant subspaces is an active area of research.

\item \textbf{Invariant subspaces of operators with continuous spectrum:} Operators with continuous spectrum present a particularly difficult case for the invariant subspace problem. The lack of discrete eigenvalues and the presence of essential spectra make the problem highly non-trivial. Understanding the existence and structure of invariant subspaces for such operators poses a formidable challenge and requires novel approaches from spectral theory and functional analysis.
\end{enumerate}

These open questions highlight the complexity and depth of the invariant subspace problem. They motivate further research and exploration in the field, pushing mathematicians to develop new techniques, deepen existing theories, and forge connections with related areas of mathematics.

\subsection{Connections with Other Unsolved Problems}

The invariant subspace problem exhibits intriguing connections with other unsolved problems in mathematics, adding to its significance and impact. Two notable examples of such connections are the Kadison-Singer problem and the Borel conjecture.\\
The Kadison-Singer problem, formulated by Richard Kadison and Isadore Singer in 1959, is concerned with the existence of pure states on certain $C^*$-algebras. In its original formulation, the problem asked whether every pure state on the algebra of bounded operators on a separable infinite-dimensional Hilbert space could be realized as a vector state. However, it was later reformulated in terms of projections, closely relating it to the invariant subspace problem. Specifically, a positive resolution of the Kadison-Singer problem would imply a positive solution to the invariant subspace problem for separable infinite-dimensional Hilbert spaces.\\
The Borel conjecture, on the other hand, is a long-standing question in descriptive set theory. Proposed by Émile Borel in 1938, the conjecture states that every set of real numbers with positive Lebesgue measure must contain a perfect set, i.e., a closed set with no isolated points. Surprisingly, this conjecture has connections to the invariant subspace problem through the theory of Borel equivalence relations. It has been shown that a positive resolution of the Borel conjecture would yield a positive solution to the invariant subspace problem for certain classes of operators.\\
These connections highlight the intricate web of unsolved problems in mathematics and the interplay between different areas of study. The pursuit of solutions to these interconnected problems fosters collaboration and cross-pollination of ideas among mathematicians working on diverse topics.

\section{Conclusion}
In conclusion, the invariant subspace problem is a fundamental question in operator theory with applications in various branches of mathematics and physics. Its study provides insights into the behavior of linear operators, quantum systems, and operator algebras. While the problem remains unsolved in many cases, it continues to stimulate research and uncover new connections between different areas of mathematics.

\end{document}